\documentclass[prd,twocolumn,floatfix]{revtex4}
\usepackage{graphicx}
\usepackage{amssymb,amsmath}
\usepackage{dcolumn}
\begin{document}

\title{Braneworlds with timelike extra-dimension}
\author{Marina Seikel}
\email{mseikel@physik.uni-bielefeld.de}
\affiliation{Fakult\"at f\"ur Physik, Universit\"at Bielefeld,
  Postfach 100131, 33501 Bielefeld, Germany}
\author{Max Camenzind}
\email{m.camenzind@lsw.uni-heidelberg.de}
\affiliation{Landessternwarte, ZAH, K\"onigstuhl 12, 69117 Heidelberg,
Germany}
\date{}

\begin{abstract}
In this work, we consider a braneworld model with a timelike
extra-dimension. There are strong constraints to the parameter values
of such a model resulting from the claim that there must be a physical
solution to the Friedmann equation at least between now and the time
of recombination. We fitted the model to supernova type Ia data and
checked the consistency of the result with other observations. For
parameter values that are consistent with observations, the braneworld
model is indistinguishable from a $\Lambda$CDM universe as far as the
considered cosmological tests are concerned.
\end{abstract}

\pacs{98.80.-k,95.36.+x,04.50.-h}

\maketitle

\section{Introduction}
After the seminal papers of Randall and Sundrum \cite{RS1,RS2} several
braneworld models 
have been examined during the past years (for reviews see
\cite{maartens04,Koyama}). The idea is that we could be living 
in a four-dimensional space-time, the brane, which is embedded in or bounding
a five-dimensional bulk. Gravity acts in all five dimensions whereas
the other interaction forces are constrained to the brane.
Thus, the existence of an additional dimension
influences the expansion history of the universe.
While the Randall-Sundrum model differs from
general relativity in the early universe, another braneworld model was
suggested by Dvali, Gabadadze and Porrati (DGP model) \cite{dvali} which
differs from general relativity at late times and thus could give an
explanation for the present accelerated expansion. In fact, braneworld
cosmologies (e.g. a combination of the Randall-Sundrum and the DGP
model \cite{collins,sahni02}) can mimick several other cosmological
models, but also 
allow for a variety of other expansion histories \cite{sahni05}.
In most models the extra-dimension is considered to
be spacelike, but there is actually no a priori reason that prevents this
dimension from being timelike. 

In the present work we consider a model with a single brane which forms
the boundary of the bulk. Chapter II summarizes the most important
properties of this model that has already been described by Shtanov
and Sahni \cite{ShtaSa,Shta}. Here, we take a look at
the general case, leaving open the nature of the fifth dimension. In
capter III we draw our attention to the case 
of a timelike extra-dimension. We put new constraints on
the density parameters in order to get a physical solution of the
Friedmann equation within a certain redshift range. At this point, the
so-called BRANE2 model can already be excluded in the case of
vanishing spatial curvature and dark radiation.
The BRANE1 model is then confronted with observational data showing
that it cannot be excluded by the applied cosmological tests.

\section{A Braneworld Model}
We consider a theory which combines the Randall-Sundrum and the DGP 
model and is described by the action \cite{ShtaSa,Shta}
\begin{eqnarray}
S &=& M^3\left[ \int_{\mbox{\footnotesize bulk}}(\mathcal{R} -
  2\Lambda_5)\sqrt{-\epsilon g}\,d^5x\right. \nonumber\\* 
  &&\qquad\left. - 2\epsilon\int_{\mbox{\footnotesize brane}}K\sqrt{-h}\,d^4x \right]
\nonumber\\*
  &&{}+ \int_{\mbox{\footnotesize brane}}(m^2R - 2\sigma)
\sqrt{-h}\,d^4x \nonumber\\* 
  &&{}+ \int_{\mbox{\footnotesize brane}}L(h_{ab},\phi)\sqrt{-h}\,d^4x \, ,
\end{eqnarray}
where $M$ and $m$ are the five- and four-dimensional Planck masses,
respectively. The two masses are related by an important length scale
$\ell=2m^2/M^3$. On short length scales $(r\ll\ell)$ the usual
four-dimensional general relativity is recovered, while on large
length scales $(r\gg\ell)$ five-dimensional effects play an important
role \cite{sahni03,dvali}. 
$\mathcal{R}$ denotes the scalar curvature of the bulk metric
$g_{ab}$ and $R$ the scalar curvature of the induced brane metric
$h_{ab}=g_{ab}-\epsilon n_an_b$, with $n^a$ being the inner unit normal vector
field to the brane. $K$ is the trace of the extrinsic curvature of the brane
$K_{ab}=h^c_{\,\,a}\nabla_cn_b$. $\Lambda_5$ denotes the bulk cosmological
constant and $\sigma$ the brane tension. As ordinary matter fields are
confined to the brane, the Lagrangian density $L$ does not depend on the bulk
metric $g_{ab}$, but on the induced metric $h_{ab}$.
For a spacelike extra-dimension $\epsilon=1$, whereas $\epsilon=-1$ for a
timelike extra-dimension.

By variation of this action one obtains Einstein's equations in the bulk
\begin{eqnarray}\label{Einstein_bulk}
\mathcal{G}_{ab} + \Lambda_5g_{ab} = 0
\end{eqnarray}
and on the brane
\begin{eqnarray}\label{Einstein_brane}
m^2G_{ab} + \sigma h_{ab} = T_{ab} + \epsilon M^3(K_{ab} - Kh_{ab})\, .
\end{eqnarray}

In order to calculate the five-dimensional Friedmann equations from the above
Einstein equations one needs the Gauss-Codacci-relation (\cite{Wald}
chapter 10.2): 
\begin{eqnarray} \label{Gauss}
R_{abc}^{\quad\;
  d}=h_a^{\;\;f}h_b^{\;\;g}h_c^{\;\;k}h^d_{\;\;j}\mathcal{R}_{fgk}^{\quad\;
  j} + K_{ac}K_b^{\;\;d} - 
K_{bc}K_a^{\;\;d}
\end{eqnarray}
Contracting this relation on the brane leads to
\begin{align}
M^6\left(R - 2\Lambda_5\right) -& \frac{1}{3}\left(m^2R - 4m^2\Lambda_4 +
 T\right)^2  \nonumber\\*
 {} +& \left(m^2G_{ab}+m^2\Lambda_4
  h_{ab}-T_{ab}\right)\nonumber\\*
 &\times\left(m^2G^{ab}+m^2\Lambda_4
  h^{ab}-T^{ab}\right) = 0
\end{align}
In the following, we consider a homogeneous and isotropic universe. Taking the
stress energy conservation into account, the above equation can be integrated
to yield \cite{Shta,deffayet}
\begin{eqnarray}\label{brane}
m^4\left(H^2 + \frac{k}{a^2} - \frac{\rho+\sigma}{3m^2}\right)^2 \nonumber\\*
= \epsilon
M^6\left(H^2 + \frac{k}{a^2} - \frac{\Lambda_5}{6} - \frac{C}{a^4}\right)\, ,
\end{eqnarray}
where $H=\dot{a}/a$ is the Hubble parameter 
and $k=0, \pm 1$ corresponds to the spatial curvature. 
$\rho$ is the matter density on the
brane. $C$ is an integration constant, the dark
radiation term, which transmits bulk graviton influence onto the brane.
Introducing the length scale $\ell=2m^2/M^3$ equation \eqref{brane}
yields the Friedmann equation on the brane
\begin{eqnarray}
&&H^2 + \frac{k}{a^2} \nonumber\\
&=& \frac{\rho + \sigma}{3m^2} \nonumber\\*
&&{}+ \epsilon
\frac{2}{\ell^2}\left[1\pm\sqrt{1 +
    \epsilon\ell^2\left(\frac{\rho+\sigma}{3m^2} -
    \frac{\Lambda_5}{6}-\frac{C}{a^4} 
   \right)}\right] \\
\label{friedmann}
&=& \frac{\Lambda_5}{6} + \frac{C}{a^4} \nonumber\\*
&&+ \epsilon\frac{1}{\ell^2} \left[1\pm
  \sqrt{1 +
    \epsilon\ell^2\left(\frac{\rho+\sigma}{3m^2}-\frac{\Lambda_5}{6}-
    \frac{C}{a^4} 
   \right)}\right]^2\, .
\end{eqnarray}
The $\pm$-sign corresponds to the two different ways the brane can be embedded
in the bulk. The model which is described by the equation with the
lower sign will from now on be 
referred to as BRANE1 and the one with the
upper sign as BRANE2.

Using the cosmological density parameters
\begin{align*}
 \Omega_m& =\frac{\rho_0}{3m^2H_0^2},\quad& \Omega_k&
  =-\frac{k}{a_0^2H_0^2},\quad& \Omega_{\sigma}& =\frac{\sigma}{3m^2H_0^2}\\
 \Omega_{\ell}& =\frac{1}{\ell^2H_0^2}, & \Omega_{\Lambda_5}&
  =-\frac{\Lambda_5}{6H_0^2}, & \Omega_C& =-\frac{C}{a_0^4 H_0^2} 
\end{align*}
the Friedmann equation can be rewritten as
\begin{eqnarray}
\frac{H^2(z)}{H_0^2} = \Omega_m(1+z)^3 + \Omega_k(1+z)^2 +
  \Omega_{\sigma}\nonumber\\*
{} + 2\epsilon\Omega_{\ell} 
{} \pm 2\epsilon\sqrt{\Omega_{\ell}}\nonumber\\* 
\times\sqrt{\Omega_\ell +\epsilon \left[\Omega_m(1+z)^3 +
  \Omega_{\sigma} + \Omega_{\Lambda_5} + \Omega_C(1+z)^4\right]}\, .
\end{eqnarray}
Considering only the first three terms on the RHS, we receive the well-known
Friedmann equation of four-dimensional general relativity. This is equivalent
to setting the five-dimensional Planck mass $M$ to zero
(i.e.~$\Omega_\ell=0$). If we instead (in the case of a spacelike
extra-dimension) choose the four-dimensional Planck mass $m$ to be
zero, the result is a Randall-Sundrum braneworld model
\cite{RS1,RS2}.

In the approach we follow in this work, there is no need to specify a
metric. The Friedmann equation could be derived without making any
assumptions on the metric except homogeneity and isotropy. This makes
the ansatz rather general. Nevertheless, we give an example for a
metric. In the case $m=0$, the bulk can be described by the following
metric \cite{ShtaSa}:
\begin{equation}
  ds^2 = -f(r)dt^2 + \frac{\epsilon dr^2}{f(r)} + r^2d\Omega_3 \;,
\end{equation}
where 
\begin{equation}
  f(r) = \epsilon\left( k - \frac{\Lambda_5 r^2}{6} - \frac{C}{r^2} \right)
\end{equation}
and $d\Omega_3$ denotes the metric of the unit three-sphere. For a
spacelike extra-dimension, this solution corresponds to
AdS$_5$-Schwarzschild.

\section{Timelike Extra-dimension}
We will now focus on the case of a timelike extra-dimension. 
Thus, the Friedmann equation is
\begin{eqnarray}
\frac{H^2(z)}{H_0^2} = \Omega_m(1+z)^3 + \Omega_k(1+z)^2 +
  \Omega_{\sigma}\nonumber\\*
{} - 2\Omega_{\ell} 
{} \mp 2\sqrt{\Omega_{\ell}}\nonumber\\* \label{fried}
\times\sqrt{\Omega_\ell - \Omega_m(1+z)^3 -
  \Omega_{\sigma} - \Omega_{\Lambda_5} - \Omega_C(1+z)^4} \, .
\end{eqnarray}
From equation \eqref{friedmann} the following constraint on the
density parameters can be obtained by setting $z=0$:
\begin{eqnarray}\label{constraint}
1 &=& \Omega_k - \Omega_{\Lambda_5} - \Omega_C \nonumber\\*
&&{}- \left[\sqrt{\Omega_\ell} \pm
  \sqrt{\Omega_\ell - \Omega_m - \Omega_\sigma - \Omega_{\Lambda_5} -
  \Omega_C}\right]^2
\end{eqnarray}
As the density parameters are real quantities,
\begin{eqnarray}
&&\Omega_\ell\ge 0,\\
&&\Omega_\ell \ge \Omega_m + \Omega_\sigma + \Omega_{\Lambda_5} + \Omega_C
\end{eqnarray}
as well as 
\begin{equation}
\Omega_k - \Omega_{\Lambda_5} - \Omega_C \ge 1
\end{equation}
must be fulfilled. For a $\Lambda$CDM universe, we know from
observations that the spatial curvature $\Omega_k$ is close to
zero. Komatsu et al.~\cite{komatsu} have constrained the spatial 
curvature to be $\Omega_k=-0.0052\pm 0.0064$ by using WMAP5 data. On the
other hand, the dark radiation density at the 
present epoch $\Omega_C$ also has to be quite small as it scales with
$(1+z)^4$. Assuming that the constraints on the spatial curvature are
also valid for braneworld models, from the above equation
follows $\Omega_{\Lambda_5} \lesssim -1$. At least $\Omega_{\Lambda_5}$ has to
be negative, i.e.~the bulk cosmological constant $\Lambda_5$ is positive.

\subsection{Vanishing Dark Radiation and Spatial Curvature}
For simplicity we first consider a model with $\Omega_k=0=\Omega_C$. 
In this
case, the only parameter that scales with redshift is the matter density. As
the BRANE1 and the BRANE2 models show quite a different behaviour, we
discuss them 
separately.

\subsubsection{BRANE1}
Two conditions have to be satisfied in \eqref{fried}: $H^2(z)$
(condition 1) as well as the 
term  under the square root (condition 2) must not be negative.\\
Condition 1:
\begin{eqnarray}
\label{cond1}
\Omega_m(1+z)^3 + \Omega_\sigma - 2\Omega_{\ell}&&\nonumber\\
{}+ 2\sqrt{\Omega_{\ell}} 
  \sqrt{\Omega_\ell - \Omega_m(1+z)^3 -\Omega_\sigma
   - \Omega_{\Lambda_5}} &\ge& 0
\end{eqnarray}
Condition 2:
\begin{eqnarray}\label{cond2}
\Omega_m(1+z)^3 \le \Omega_\ell - \Omega_{\Lambda_5} - \Omega_\sigma\, .
\end{eqnarray}

From the constraint equation \eqref{constraint} we receive two solutions for
the brane tension 
\begin{equation}
\Omega_\sigma = 1 - \Omega_m \pm
2\sqrt{\Omega_\ell}\sqrt{-1-\Omega_{\Lambda_5}} \,.
\end{equation}
The constraints on the density parameters given by condition 1 and 2
strongly depend on whether a negative or positive brane tension is
chosen. 

In the case of a {\em negative} $\Omega_\sigma$, either the inequality
\begin{equation}
\Omega_\ell \ge \left( \frac{\Omega_m(1+z)^3+1-\Omega_m}{2\left(
      \sqrt{-\Omega_{\Lambda_5}} + \sqrt{-\Omega_{\Lambda_5}-1} \right)} \right)^2
\end{equation}
or the inequalities
\begin{eqnarray}
\sqrt{\Omega_\ell} &\ge& \sqrt{\Omega_m(1+z)^3-\Omega_m} -
\sqrt{-1-\Omega_{\Lambda_5}} \quad \text{and} \nonumber\\ 
\Omega_\ell &\le& -\Omega_{\Lambda_5}
\end{eqnarray}
have to be fulfilled.

For a {\em positive} $\Omega_\sigma$, 
\begin{equation}
\Omega_\ell \ge \left( \frac{\Omega_m(1+z)^3+1-\Omega_m}{2\left(
      \sqrt{-\Omega_{\Lambda_5}} - \sqrt{-\Omega_{\Lambda_5}-1} \right)} \right)^2
\end{equation}
must be fulfilled.

From the constraint inequalities one can immediately see that there
exists a maximum redshift beyond which the Friedmann equation does not
have a physical solution. We claim that the conditions must at least
be satisfied back to the time of recombination, i.e. $z=1090$. When
going to higher redshifts, at some point either \eqref{cond1} or
\eqref{cond2} is violated. The physical meaning of the former case is
the following: In a collapsing universe the Hubble parameter
$H(z)$ becomes zero at a certain redshift and a bounce takes place
\cite{ShtaSa}. After that the universe expands again. In contrast,
violation of \eqref{cond2} would lead to a singularity similar to those
described in \cite{sing}: The deceleration parameter
$q(z)=-\ddot{a}/(aH^2)$ becomes singular whereas $H(z)$ remains finite.

\begin{figure}
\includegraphics{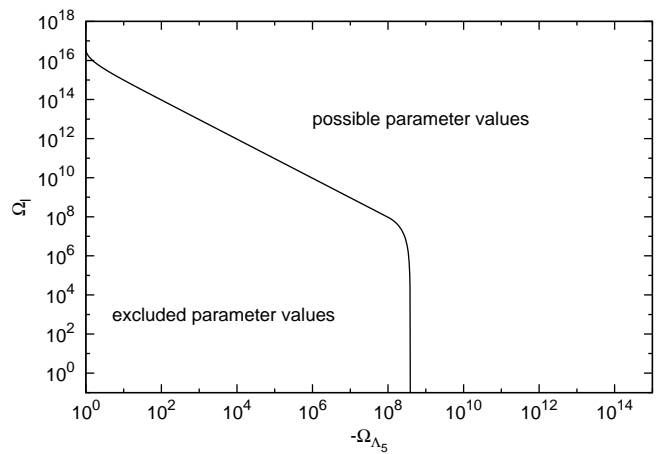}
\caption{\label{ol5_ol} Constraints on the density parameters $\Omega_\ell$
  and $\Omega_{\Lambda_5}$ of a BRANE1 model with {\em negative} brane
  tension and $\Omega_m=0.3$.}
\end{figure}

\begin{figure}
\includegraphics{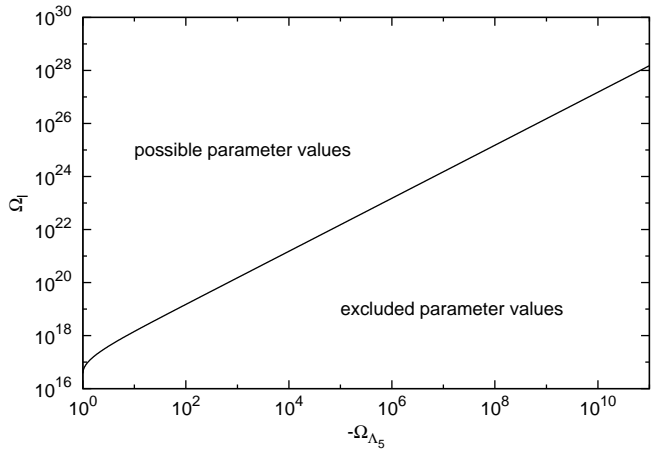}
\caption{\label{pos-os} Constraints on the density parameters $\Omega_\ell$
  and $\Omega_{\Lambda_5}$ of a BRANE1 model with {\em positive} brane
  tension and $\Omega_m=0.3$.}
\end{figure}

The conditions that constrain the density parameters in the cases of
negative and positive brane tension are quite similar. 
Yet, the consequences for the allowed parameter space are
very different as can be seen in Figs.~\ref{ol5_ol} and \ref{pos-os}.
The range of possible parameter values is much larger for a negative
brane tension. Therefore, we will focus on this case in the following.


\begin{figure}
\includegraphics{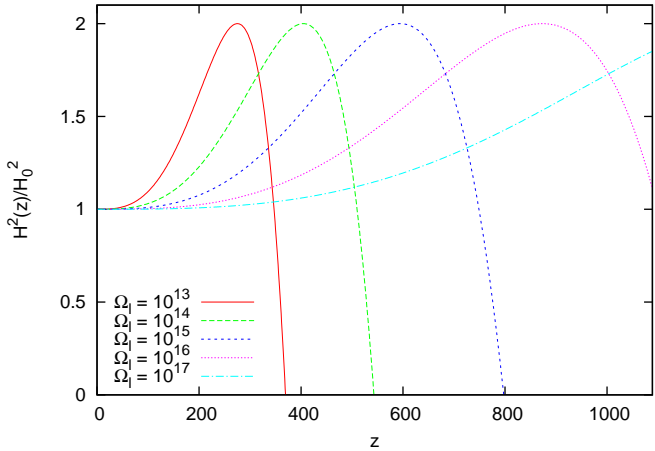}
\caption{\label{hubble} Hubble parameter $H^2(z)/H^2_0$ of a BRANE1 model
  with negative brane tension for different values
  of $\Omega_\ell$ with $\Omega_m=0.3$ and $\Omega_{\Lambda_5}=-2$.}
\end{figure}

\begin{figure}
\includegraphics{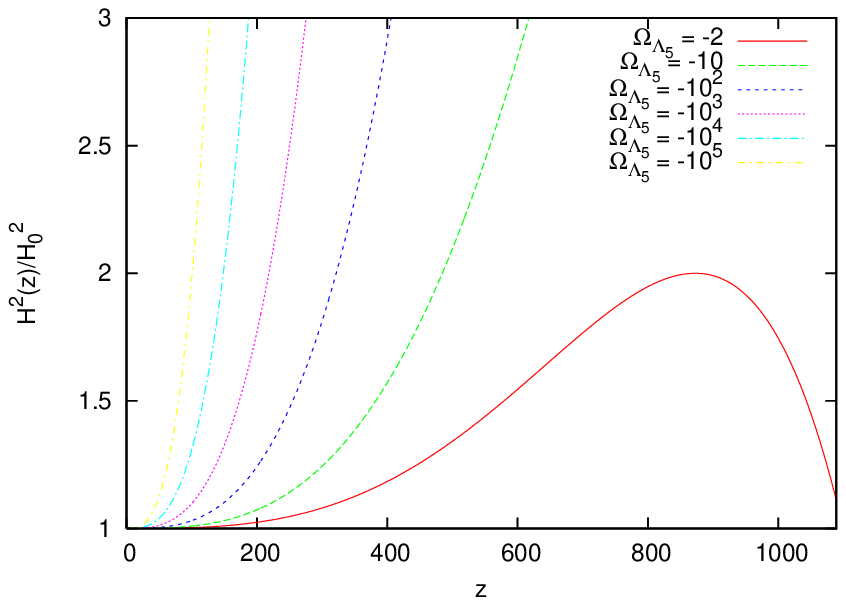}
\caption{\label{hubble_l5} Hubble parameter $H^2(z)/H^2_0$ of a BRANE1 model
  with negative brane tension for different values
  of $\Omega_{\Lambda_5}$ with $\Omega_m=0.3$ and $\Omega_\ell=10^{16}$.}
\end{figure}

Figure \ref{hubble} shows how the square of the
Hubble parameter changes with increasing $\Omega_\ell$, while
$\Omega_m=0.3$ and $\Omega_{\Lambda_5}=-2$ stay fixed for the case of a
negative brane tension. For values of $\Omega_\ell$ smaller than
$\sim 10^{16}$ (i.e. values within the excluded parameter range), the
curve becomes negative between now and the time of recombination. In
Fig.~\ref{hubble_l5}, it is shown how the curve changes when
$\Omega_{\Lambda_5}$ is varied. Here, all parameter values are in the
allowed range. With increasing $\Omega_{\Lambda_5}$ the curves become
steeper and thus approach the $\Lambda$CDM model (which cannot be
shown in this figure as it is too steep).

\subsubsection{BRANE2}
For the BRANE2 model again two conditions have to be fulfilled:
\begin{eqnarray}
\Omega_m(1+z)^3 + \Omega_\sigma - 2\Omega_{\ell} &&\nonumber\\
{}- 2\sqrt{\Omega_{\ell}} 
  \sqrt{\Omega_\ell - \Omega_m(1+z)^3 -\Omega_\sigma
   - \Omega_{\Lambda_5}} &\ge& 0
\label{condition1}
\end{eqnarray}
and
\begin{eqnarray}
\Omega_\ell - \Omega_m(1+z)^3 -\Omega_\sigma
   - \Omega_{\Lambda_5} \ge 0 \,,
\label{condition2}
\end{eqnarray}
where the brane tension is given by $\Omega_\sigma = 1 - \Omega_m +
2\sqrt{\Omega_{\ell}}\sqrt{-1-\Omega_{\Lambda_5}} $.

These conditions are equivalent to the inequalities
\begin{equation}\label{B2constraint1}
\Omega_{\ell} \le
\frac{1}{4(\sqrt{-\Omega_{\Lambda_5}}-\sqrt{-1-\Omega_{\Lambda_5}})^2} 
\end{equation}
and
\begin{equation}\label{B2constraint2}
\Omega_\ell \ge (\sqrt{\Omega_m(1+z)^3-\Omega_m} +
\sqrt{-1-\Omega_{\Lambda_5}})^2 \,.
\end{equation}
The two inequalities can only be fulfilled simultaneously, if the RHS of
\eqref{B2constraint1} is larger than the RHS of
\eqref{B2constraint2}. This is only possible for redshifts $z\lesssim
0.22$. Thus, a flat BRANE2 model without dark radiation can be excluded.

\subsubsection{Angular Separation}
In the following we will concentrate on the BRANE1 model with negative
brane tension. Remember that in this section we consider the universe
to be spatially flat and to contain no dark radiation.
We would now like to know whether such a model is compatible with
observations. One simple cosmological test is to take a look at
the angular separation. The angle $\Theta(z)$ under which we see two
objects in the universe depends on the cosmological model. As the
universe expands, the distance $D(z)$ between those objects
changes as
\begin{equation}
D(z) = \frac{D_0}{1+z}\, ,
\end{equation}
where $D_0=D(z=0)$ is the separation in the present universe.
The angular separation is described by
\begin{eqnarray}
\Theta (z) &=& \frac{D(z)(1+z)^2}{d_L(z)} = \frac{D_0(1+z)}{d_L(z)} \nonumber\\*
&=& D_0\left[\int_0^z\frac{dz'}{H(z')}\right]^{-1} \, ,
\end{eqnarray}
where $d_L(z)$ is the luminosity distance and the last equation is only valid
for a flat universe, $\Omega_k=0$.

\begin{figure}
\includegraphics{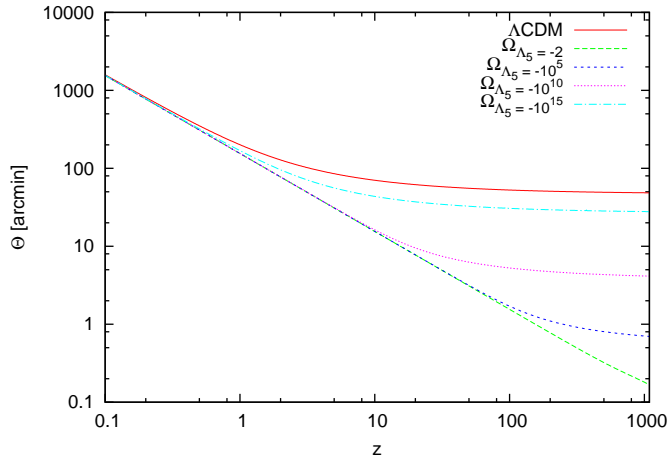}
\caption{\label{ang} Angular separation for a $\Lambda$CDM ($\Omega_m=0.3$,
  $\Omega_\Lambda=0.7$) and a BRANE1 model for different
  values of $\Omega_{\Lambda_5}$ 
with $\Omega_m=0.3$, $\Omega_\ell=10^{16}$ and $D_0=185$Mpc.}
\end{figure}

The large-scale correlation function of luminous red galaxies has been
obtained from Sloan Digital Sky Survey data showing a peak at 100
$h^{-1}$ Mpc \cite{eisenstein}. The average redshift of those galaxies
is $z=0.35$. Assuming $h=0.73$, we can determine the present
separation to be $D_0=185$ Mpc. If one uses this typical distance of large
objects in the present universe and calculates $\Theta$ for $z=1090$
with the above formula, the resulting angle should be a typical value
for the structure observed in the CMB. Figure \ref{ang} shows the
angular separation for $\Lambda$CDM and for BRANE1. 
$\Theta(z=1090)=48.5$arcmin in a $\Lambda$CDM universe, which is where
the first peak of the CMB power spectrum is located. Thus,
$\Lambda$CDM fits the observational data perfectly well.
For small absolute values of $\Omega_{\Lambda_5}$ in the braneworld
model, the angular width $\Theta(z=1090)$ is about 300 times 
smaller than in the $\Lambda$CDM case and thus not compatible with CMB
observations. This cannot be remedied by changing the value of $\Omega_\ell$.
The larger $\Omega_{\Lambda_5}$ is chosen the more the
angular width approaches that of $\Lambda$CDM.
Therefore a small $\Omega_{\Lambda_5}$ can be ruled out for
this special kind of braneworld model with $\Omega_k=0$ and $\Omega_C=0$.

\subsection{BRANE1 with Dark Radiation and Spatial Curvature}
In this section we give up the assumption of a flat universe without dark
radiation. Instead we assume a negative $\Omega_C$,
which corresponds to a positive dark radiation term $C$. $\Omega_k$
can have arbitrary values.

\begin{figure}
\includegraphics{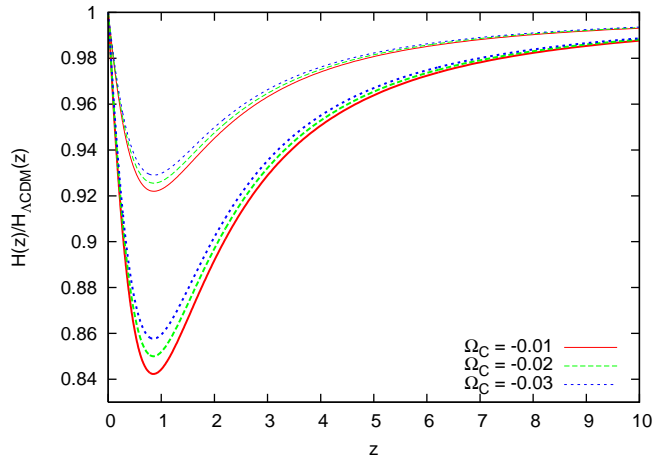}
\caption{\label{hubble_c} $H(z)/H_{\Lambda\mbox{\footnotesize CDM}}(z)$ for a
  BRANE1 closed universe with dark radiation. $\Omega_m=0.3$ and
  $\Omega_k=-0.01$ for all curves, $\Omega_{\Lambda_5}=-2$ and
  $\Omega_\ell=10^{16}$ for the lower three bold lines and
  $\Omega_{\Lambda_5}=-10^8$ and  $\Omega_\ell=10^8$ for the upper
  three lines.}
\end{figure}

Figure \ref{hubble_c} shows $H(z)/H_{\Lambda\mbox{\footnotesize
    CDM}}(z)$ for different parameter values. Without  
dark energy (and with small $\Omega_{\Lambda_5}$) there was a large difference
between the braneworld model and 
$\Lambda$CDM. $H^2(z)$ became even negative at a certain redshift. Models
including dark radiation
only deviate from the standard model at relatively low
redshifts. The largest deviations occur around redshift $\simeq 1$. With
increasing $z$ the Hubble parameter approaches that of the
$\Lambda$CDM case.

A good method to compare the predictions for those redshifts with observations
is to consider the luminosity distance $d_L(z)$ or the distance
modulus $\mu(z)$. 
The luminosity distance is given by 
\begin{equation}
d_L(z) = \frac{1+z}{H_0\sqrt{|\Omega_k|}} \,\mathcal{S} \left( \sqrt{|\Omega_k|} \int_0^z
  \frac{H_0\, dz'}{H(z')} \right) \, ,
\end{equation}
where $\mathcal{S}(x)=x$ for a flat, $\sin(x)$ for a closed and
$\sinh(x)$ for an open universe.
The distance modulus is defined as
\begin{equation}
\mu(z) = m(z) - M = 5\log d_L(z) + 25 \, ,
\end{equation}
where $d_L$ is given in units of Mpc. $m(z)$
and $M$ are the apparent and the absolute magnitude, respectively. 
The observational data are obtained by analyzing
supernovae type Ia as they are considered to be the best standard candles.

We used the 2007 Gold sample presented by Riess et al.~\cite{riess07} to
fit the model. We adopted the values of $M$ and $H_0$ given by
\cite{riess05}. Therefore, we had to substract 0.27mag from distance
modulus given in the Gold sample. Then the value of $H_0$ is 73km/(s Mpc).
The problem with the $\chi^2$-fit is that there exist multiple local
minima for $\chi^2$ with many of those minima having the same value of
$\chi^2$. As shown in Fig.~\ref{hubble}, for some parameter values
$H^2(z)$ becomes zero before $z=1090$ is reached. Fits that yielded such
values could be dismissed at once. Fitting all five parameters ($\Omega_m$,
$\Omega_\ell$, $\Omega_{\Lambda_5}$, $\Omega_C$ and $\Omega_k$), the results
for $\Omega_k$ were always negative and typically between $-0.2$ and
$-0.6$. An example of best fit 
parameters is given in table \ref{fittab}. If we accept the
results from the WMAP observation (that were obtained by assuming
$\Lambda$CDM) to be valid also for braneworld models, then the result
of our fit is not compatible with WMAP which predicts a flat
universe \cite{komatsu}. Also the calculated value for $\Omega_C$ is
quite surprising. As the dark radiation density scales with $(1+z)^4$, it
should be close to zero at the present epoch. So if the value $-1.15$ is
correct, $\Omega_C$ must have been extremely large at earlier
times. 
Yet, it is noticeable that the braneworld
model with a $\chi^2$ per degree of freedom of 0.89 fits the supernova
data slightly better than $\Lambda$CDM with a $\chi^2$ per degree of
freedom of 0.90.

\begin{table}
\begin{ruledtabular}
\begin{tabular}{ccccr@{}lr@{}lc}
& $\Omega_m$ & $\Omega_\ell$ & $\Omega_{\Lambda_5}$ & 
\multicolumn{2}{c}{$\Omega_C$} & \multicolumn{2}{c}{$\Omega_k$} &
$\chi^2_{\mbox{\scriptsize dof}}$ \\
\hline
5-parameter & 0.38 & $1.9\cdot 10^{14}$ & $-7.8\cdot 10^{16}$ & $-1$&$.15$
& $-0$&$.45$ & 0.89 \\
4-parameter & 0.27 & $1.7\cdot 10^{15}$ & $-6.0\cdot 10^{13}$ &
$-0$&$.17$ & $0$ && 0.89 \\
3-parameter & 0.31 & $2.0\cdot 10^{14}$ & $-1.5\cdot 10^{16}$ &0& &
$0$ && 0.91\\
\vspace{-0.7em}\\
$\Lambda$CDM & 0.28 && $\Omega_\Lambda=0.72$ &&&&& 0.90
\end{tabular}
\end{ruledtabular}
\caption{\label{fittab} Results of the $\chi^2$-fits for 3-, 4- and
  5-parameter fits of the BRANE1 model and for $\Lambda$CDM.}
\end{table}

We fixed $\Omega_k$ to be equal to zero and performed a 4-parameter fit.
The $\chi^2$ per degree of freedom for that fit is 0.89. This is
the same value as for the 5-parameter fit.
The absolute value of $\Omega_C$ has become smaller compared to the
previous fit. But still it seems to be quite large.
Performing another fit with $\Omega_C$ and $\Omega_k$ fixed to zero 
yields a $\chi^2$ per degree of freedom of 0.91, which is
slightly worse than that of $\Lambda$CDM. The density parameters have
reasonable values.

\begin{figure}
\includegraphics{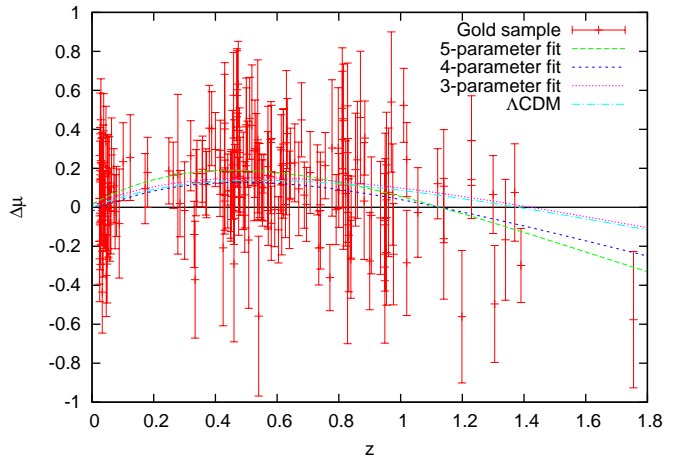}
\caption{\label{redmod} Distance modulus minus the distance modulus of
  an empty universe for the three braneworld model fits and $\Lambda$CDM.} 
\end{figure}

\begin{figure}
\includegraphics{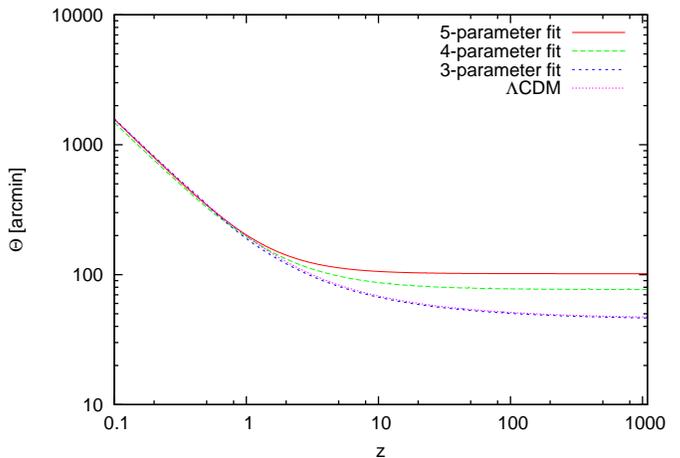}
\caption{\label{anglefit} Angular separation for the three braneworld model
  fits and $\Lambda$CDM.} 
\end{figure}

Figure~\ref{redmod} shows the distance modulus for the three fits and the
$\Lambda$CDM fit compared to the Gold sample. 
In Fig.~\ref{anglefit} the angular separation of the same models is
plotted. In both plots the curve of the 3-parameter fit is almost
identical to that of $\Lambda$CDM. While the 4- and 5-parameter fits
are perfectly consistent with SNe observations, the calculated angular
separations at redshift 1090 are too large to be compatible with CMB
observations, namely $\Theta\simeq80$ arcmin for the 4-parameter fit and
$\Theta\simeq100$ arcmin for the 5-parameter fit. Thus, the results obtained by
those two fits to SN data can be ruled out and we are left with the
result of the 3-parameter fit. 
This model almost does not differ from $\Lambda$CDM as far as the
distance modulus and the angular separation are concerned and
thus both theories are indistinguishable when using only the two
applied test. 

Table \ref{fittab2} lists the following quantities for the three
fits: a) the angular separation $\Theta$ at the time recombination, b)
the maximum possible redshift $z_{\text{max}}$ for which the Friedmann
equation has a physical solution and c) the age of the universe. For
the calculation of the maximum redshift and the age of the universe,
one needs to consider the radiation density $\Omega_r$ in the
Friedmann equation, which could be neglected in the previous tests,
but becomes important at very high redshifts. In order to do so, we
just need to add $\Omega_r(1+z)^4$ to $\Omega_m(1+z)^3$ every time it
occurs in the Friedmann equation. We adopt the value
$\Omega_r = 8.4\cdot 10^{-5}$ according to WMAP5 \cite{komatsu}. 
The redshift is only limited for the 3-parameter fit, where the term
under the square root in the Friedmann equation \eqref{fried} becomes
zero at $z_{\text{max}}$. At this point a singularity occurs, which
can be interpreted as a kind of Big Bang. 

\begin{table}
\begin{ruledtabular}
\begin{tabular}{cccc}
& $\Theta$[arcmin]  & $z_{\text{max}}$ & age [Gyr] \\
\hline
5-parameter fit & 104 & $\infty$ & 10.4 \\
4-parameter fit &  77 & $\infty$ & 12.9 \\
3-parameter fit &  46 & 120000 & 15.0
\end{tabular}
\end{ruledtabular}
\caption{\label{fittab2} Angular separation $\Theta$ at recombination,
maximum redshift $z_{\text{max}}$ and age of the universe for the
different fits.}
\end{table}

Let us take a closer look at the result of the 3-parameter fit since
this is the only model that has not been excluded by the tests used in
this work. In this model, the universe would be 15 billion years old,
i.e. there is no conflict with the oldest objects in the universe. It
has already been pointed out in \cite{sahni03} that the usual four
dimensional general relativity is recovered on scales much smaller
than $\ell$. Taking the fit result $\Omega_\ell = 2 \cdot 10^{14}$ and
assuming $H_0=73$ km/(s Mpc), one obtains $\ell \simeq 300$ pc. Thus, the
model is not in conflict with any tests of general relativity on
scales much smaller than 300 pc. Especially tests that are made within
the solar system are not affected by any five-dimensional effects.

A problem occurs when we consider Big Bang nucleosynthesis (BBN). The
maximum redshift of the 3-parameter fit is much smaller than the
redshift when nucleosynthesis took place. This problem can, however,
be easily avoided by introducing again a dark radiation term
$\Omega_C$. Its value must be small enough to ensure that the fit
result and the cosmological tests up to the redshift of recombination
are not affected. On the other hand, $\Omega_C$ needs to be larger
than the radiation density $\Omega_r$ to prevent the term under the
square root of the Friedmann equation from becoming negative. Thus, we
choose $\Omega_r$ to be of order $10^{-4}$. Then the
model is radiation dominated at very high redshifts, just like the
$\Lambda$CDM model. There is no limit to the redshift any more and the
model is consistent with BBN observations as it does not differ from
$\Lambda$CDM at these redshifts. 
Thus, this model cannot be excluded by the considered observations.

Remember that these results are only examples as the $\chi^2$-fit
yields many minima. However, we did not find a result of the 5- or
4-parameter fit that is compatible with all observations.

\section{Conclusion}
In this work we focused on braneworld models with a timelike
extra-dimension. For a flat universe without dark radiation we put
constraints on the density parameters $\Omega_{\Lambda_5}$ and
$\Omega_\ell$. The BRANE2 model could be excluded for this
case. Considering a BRANE1 model, the absolute value of at least one
of the parameters $\Omega_{\Lambda_5}$, $\Omega_\ell$ has to be very large
in order to obtain a physical solution for the Friedmann equation
within a redshift range from 0 to 
1090. Comparison to CMB data shows that a large
$|\Omega_{\Lambda_5}|$ is necessary for this model. 

We then introduced a dark radiation term and spatial curvature and
fitted the density parameters to SN Ia data. The results of the 5- and
4-parameter fits are not compatible with CMB observations. The only
result that could not be ruled out is the 3-parameter fit, provided a
small dark radiation term is present. Unfortunately, its behaviour in
the considered cosmological tests
is almost identical to that of $\Lambda$CDM. So, better observational
data would not help excluding or confirming the model. Instead, further
cosmological tests are needed. 

\begin{acknowledgments}
We thank Dominik J. Schwarz for useful discussions and comments.
The work of MS is supported by the DFG under grant GRK 881.
\end{acknowledgments}

\bibliography{brane}

\end{document}